\documentclass[%
reprint,
 amsmath,amssymb,
aps,
prb,
]{revtex4-1}

\usepackage[final]{graphicx}
\usepackage{dcolumn}
\usepackage{float}
\usepackage{subcaption}
\usepackage{caption}
\usepackage{csquotes}

\usepackage{changepage}
\usepackage{physics}
\usepackage{color,soul}
\usepackage{bm}

\newcommand{\er}{Er$_2$Ti$_2$O$_7$}
\newcommand{\yb}{Yb$_2$Ti$_2$O$_7$}

\newcommand{\nd}{Nd$_2$Zr$_2$O$_7$}
\newcommand{\ndsn}{Nd$_2$Sn$_2$O$_7$}
\newcommand{\ndhf}{Nd$_2$Hf$_2$O$_7$}
\newcommand{\ndir}{Nd$_2$Ir$_2$O$_7$}
\newcommand{\liga}{LiGaCr$_4$O$_8$}
\newcommand{\liIn}{LiInCr$_4$O$_8$}
\newcommand{\bayb}{Ba$_3$Yb$_2$Zn$_5$O$_{11}$}

\begin{document}

\preprint{APS/123-QED}

\title{Analytic description of spin waves in dipolar/octupolar pyrochlore magnets}

\author{I. A. Assi}
\email{iassi@mun.ca}
\author{S. H. Curnoe}%
\affiliation{Department of Physics and Physical Oceanography, Memorial University of Newfoundland, St. John’s, Newfoundland \& Labrador, A1B 3X7, Canada.
}
\date{\today}
\begin{abstract}
 We derive analytic forms for spin waves in pyrochlore magnets with dipolar-octupolar interactions, such as \nd.  We obtain full knowledge of the diagonalized magnonic Hamiltonian within the linear spin wave approximation. We also  consider the effect of a ``breathing mode" as a  perturbation of this system. The breathing mode lifts the degeneracy of the upper band of the spin wave dispersion along the direction $X\to W$ in  $k$-space. 
\end{abstract}

\maketitle

\section{Introduction}

Rare earth pyrochlore oxides 
are systems with chemical formula ${\rm A_{2}B_{2}O_{7}}$, where ${\rm A}$ is a rare earth (RE) ion, and ${\rm B}$ refers to a transition metal ion. 
Both the rare-earth the transition metal ions are arranged on lattices which are corner-sharing tetrahedral
networks, an arrangement that may lead to geometrical frustration.
Depending on the choice of the RE, these compounds 
show a range of interesting 
states at low temperature, including spin ice  in 
${\rm Ho_{2}Ti_{2}O_{7}}$, and ${\rm Dy_{2}Ti_{2}O_{7}}$,
\cite{harris1997, ramirez1999} a quantum spin liquid  state in ${\rm Tb_{2}Ti_{2}O_{7}}$, \cite{gardner2001} and antiferromagnetic ordering in ${\rm Er_{2}Ti_{2}O_{7}}$\cite{Er1champion2003} and $\rm Nd_{2}Zr_2O_7$. \cite{lhotel2015,  Nd-xu2019} 

Magnetic order arises from interactions between the rare earth spins. The most general, symmetry-allowed form of the exchange interaction on the pyrochlore lattice has four independent exchange constants,\cite{curnoe2008} and much effort has been spent
to determine these constants for different pyrochlore crystals and to find the phase diagram for this four-parameter space.
Recently, Yan {\em et al.}\cite{yan2017} have determined the phase space that encompasses different kinds
of magnetically ordered states. 
The exchange interaction can also account for 
excitations (i.e., magnons) above the magnetically ordered ground state. In fact, the measurement of the 
dispersion relation of magnons in \yb\ and \er\ was used to determine the value of the exchange 
constants for those materials.\cite{ross2011, Er3savary2012}

In pyrochlores, the crystal electric field (CEF) at the rare site lifts the $2J+1$-fold degeneracy of 
the rare-earth spin $J$ into singlets and doublets. 
The CEF states are associated with irreducible representations of $D_{3}'$, the point group symmetry of  the CEF.
For integer $J$, there are $\Gamma_1$ or $\Gamma_2$ singlet states, as well as non-Kramers doublets,
$\Gamma_3$.
For half-integer $J$ there are two kinds of Kramers doublets, $\Gamma_4$ (spin-1/2) and $\Gamma_{5,6}$ (dipolar/octupolar). When the 
energy difference between the ground state  and the first excited state is large (of the order of 100 K) one can neglect all of the CEF levels except the lowest. When the CEF ground state is a doublet, the result is a frustrated lattice of interacting two-state systems which may be treated as pseudo-spins. General forms of nearest-neighbour interactions for all three kinds of pseudo-spin doublets have been found.\cite{curnoe2008, huang_quantum_2014, Onoda_PhysRevLett.105.047201}

In this work, we consider systems that have
a dipolar-octupolar ($\Gamma_{5,6}$) CEF ground state doublet.
Several of these systems order in an ``all-in-all-out" (AIAO) magnetically ordered state - a state where the spins on each  tetrahedron alternate between configurations in which they all either point {\em toward} the tetrahedron centre, or {\em away} from it, including
\ndsn, which orders below $T_C = 0.91$ K,
\cite{Ndsn-bertin2015} \ndhf, with $T_C = 0.55$ K,\cite{Ndhf-anand2015} \ndir\ 
with $T_C = 15$ K,\cite{Ndir-tomiyasu2012} and \nd\ which enters this magnetic state below 0.285 K.\cite{lhotel2015} In this work, we will consider \nd\ as an illustrative example as its exchange constants are  known.\cite{Nd-petit2016,benton2016,Nd-lhotel2018,Nd-xu2019}

\begin{figure}[ht]
\includegraphics[width=0.35\textwidth]{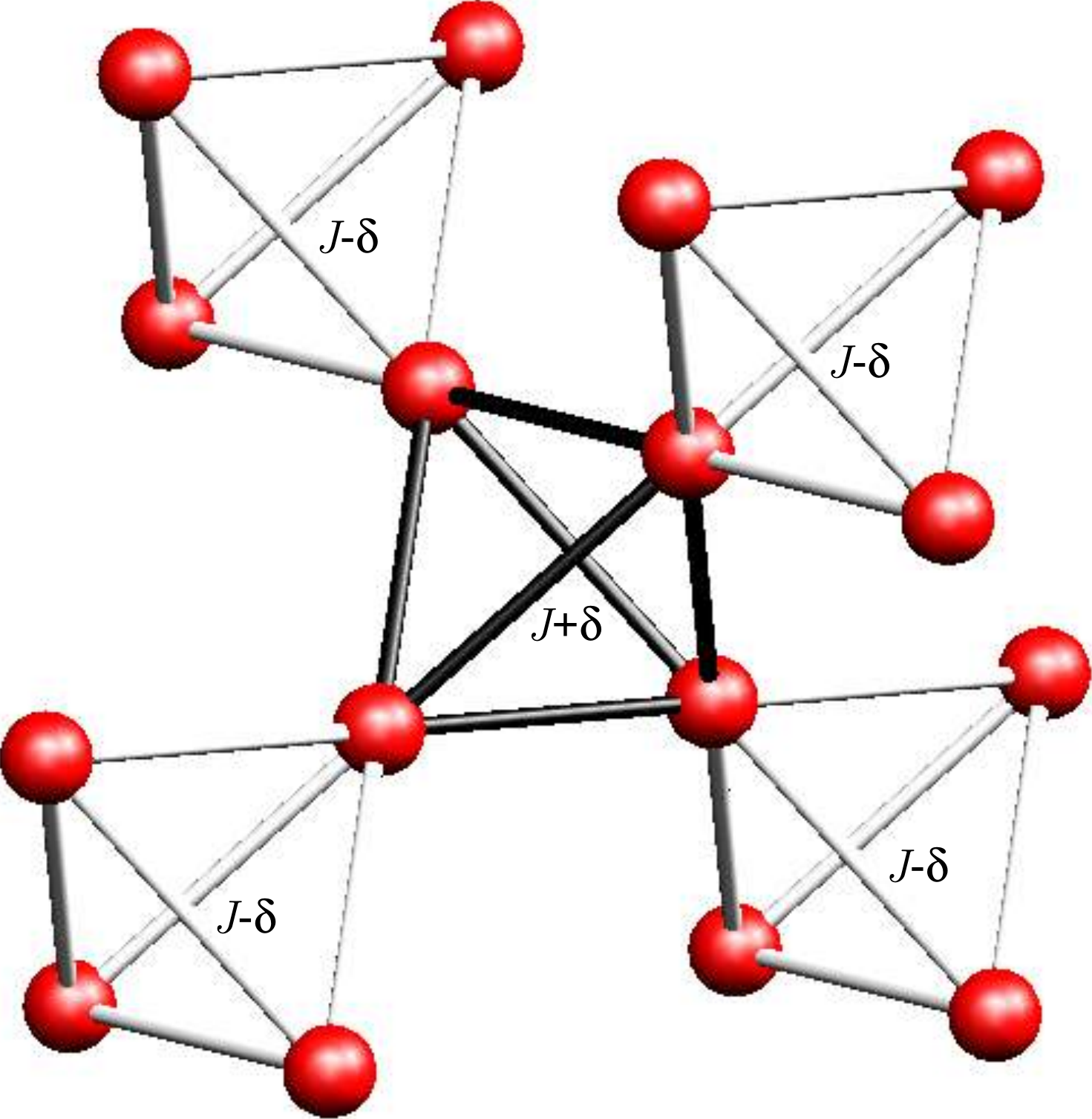}
\caption{The pyrochlore structure. The tetrahedra alternate between two orientations (black/white).
In the presence of a breathing mode the 
coupling constants associated with each exchange path will differ depending on which tetrahedron orientation they belong to.}
    \label{fig:ABtetra}
\end{figure}

We will also consider the effect of a \textit{breathing mode} on the pyrochlore lattice.
The phenomena of a breathing pyrochlore lattice was first realized in spinel oxides \liga\ and \liIn\ in which alternating tetrahedra expand and contract.\cite{Lispinels} Other materials that exhibit the breathing mode include \bayb\cite{bayb1,bayb3} and chromium spinel sulfides.\cite{chrospinels} In recent years the breathing mode has been explored theoretically and experimentally in many different contexts.\cite{curnoe2008,benton-ground2015,li_weyl_2016,savary_quantum_2016, rau_anisotropic_2016,essafi_flat_2017,jian_weyl_2018,ezawa_higher-order_2018,aoyama-spin2019,hirschberger_skyrmion_2019, talanov_formation_2020,HanYanBPy2020,Wakao2020,Masaki2020,Reschke2020, shahzad2020}
The breathing mode can be parameterized in terms of a \textit{breathing factor} which is the ratio between the exchange constants on the alternating tetrahedra (see Fig.\ \ref{fig:ABtetra}).  In the general anisotropic exchange model, for each independent exchange constant there is an independent breathing factor.\cite{curnoe2008}
            
In the following, we present {\em analytic} calculations of magnon dispersions for the $\Gamma_{5,6}$ ground state doublet in the presence of an AIAO magnetic state. A bosonic Hamiltonian describing magnons is obtained from the exchange  Hamiltonian using the Holstein-Primakoff transformation.\cite{HoPr,benton2016}
The quadratic bosonic Hamiltonian is exactly diagonalizable for all $\vec{k}$, {\em i.e.} the dispersion and the Bogoliubov transformation are presented in  analytic form. We find that the breathing mode lifts the degeneracy which otherwise occurs in the upper band along the path between the $X$-point and the $W$-point in $\vec{k}$-space.Lastly, we compute the dynamical structure factor for inelastic neutron scattering.

\section{Spin Hamiltonian}

Considering short-range interactions only, the
 Hamiltonian for the rare-earth spins has three contributions, the crystal electric field ${\mathcal H}_{\rm CEF}$, the
nearest-neighbour exchange interaction ${\mathcal H}_{\rm ex}$, and the Zeeman term ${\mathcal H}_{\rm Z}$,
    \begin{equation}
\label{eqn:totalH}
    {\mathcal H}={\mathcal H}_{\rm CEF}+{\mathcal H}_{\rm ex}+{\mathcal H}_{\rm Z}.
\end{equation}
For the rare-earth site symmetry $D_{3d}$, there are six
independent terms in ${\mathcal H}_{\rm CEF}$ which are expressed as Stevens operators.\cite{stevens1951}
The six CEF parameters have been determined (via
inelastic neutron scattering experiments) for many compounds in the pyrochlore family.

The energy scale of ${\mathcal H}_{\rm CEF}$ is higher than the other terms in ${\mathcal H}$, 
so to lowest order in perturbation theory, we consider the restriction of ${\mathcal H}$
to the degenerate ground state of ${\mathcal H}_{\rm CEF}$. The result has different 
forms depending on the symmetry of the CEF  of ground state.\cite{curnoe2017}For the $\Gamma_{5,6}$ CEF ground state the restricted exchange Hamiltonian takes the  general form\cite{huang_quantum_2014}
\begin{eqnarray}
    {\mathcal H}^{\Gamma_{5,6}}_{\rm ex} & = &\sum_{\langle ij\rangle}[J_{z}S_{iz}S_{jz}+J_{y}S_{iy}S_{jy}+J_{x}S_{ix}S_{jx} \nonumber \\
& & +J_{zx}(S_{iz}S_{jx}+S_{ix}S_{jz})]
\label{eqn:Gamma56}
\end{eqnarray}
where $J_{z}$,$J_{y}$, $J_{x}$ and $J_{zx}$ are exchange constants; the sum is over pairs of nearest-neighbour spins and the subscripts $x, y, z$ refer to 
{\em local} coordinates where the local 
$z$ axis points along the 3-fold axis of the CEF (see Appendix \ref{appendix:a}). The pseudospin operator $\vec{S}$ acts within the 
space of the ground state doublet, and is
represented by the $2\times 2$ Pauli matrices, $\vec{S} = \frac{1}{2} \vec{\sigma}$.
The last term in Eq.\ \ref{eqn:Gamma56} can be eliminated by a rotation by an angle 
$\theta = \frac{1}{2} \tan^{-1}\left(\frac{J_{zx}}{J_x-J_y}\right)$ about the local $y$-axis, yielding\cite{huang_quantum_2014}
\begin{equation}
\label{eqn:Gamma56eff}
    {\mathcal H}_{\rm ex}^{\Gamma_{5,6}}=\sum_{\langle ij\rangle}\left[\tilde{J}_{z}\tilde{S}_{iz}\tilde{S}_{jz}+\tilde{J}_{y}\tilde{S}_{iy}\tilde{S}_{jy}+\tilde{J}_{x}\tilde{S}_{ix}\tilde{S}_{jx}\right]
\end{equation}
where $\tilde{S}_{x}$ and $\tilde{S}_{z}$ are rotated operators
and
$\tilde{J}_{x}$ and 
$  \tilde{J}_{z}$ are renormalized exchange constants resulting from the rotation, which are related to the original constants by
$J_x = \tilde{J}_x \cos^2\theta + \tilde J_{z} \sin^2\theta$ and
$J_z = \tilde{J}_x \sin^2\theta + \tilde J_{z} \cos^2\theta$. \cite{huang_quantum_2014,benton2016} We also take $\tilde{J}_y \equiv J_y$ and $\tilde{S}_y \equiv S_y$.
For \nd, the value of $\theta = 0.98$ rad was recently reported, and the exchange constants are 
  $\tilde{J}_{x}=0.091$ meV, $\tilde J_{y}=0.014$ meV, and $\tilde{J}_{z}=-0.046$ meV.\cite{Nd-xu2019}

The pseudospin operators $S_{i\alpha}$ correspond to different dipolar and octupolar {\em physical} operators that act on the space of the ground state doublet, in particular $\hat{J}_{z} = j S_z + t S_x$, where $j$ and $t$ are material dependent parameters that can be computed given an explicit form of the CEF doublet. 
For example, the CEF ground state for  \nd\ is\cite{lhotel2015}
\begin{equation}
    \ket{\pm} =  -.878\ket{\mp 9/2}\mp .05 \ket{\mp 3/2}+.476\ket{\pm 3/2}\pm .009\ket{\pm 9/2}
\end{equation}
from which we compute $j= \langle+|\hat{J}_z|+\rangle = -3.13$, and $t= \langle +|\hat{J}_z|-\rangle = 0.000282$.

For the breathing pyrochlores, we  consider the Hamiltonian (\ref{eqn:Gamma56eff}) with the 
exchange constants replaced by $\tilde{J}_{\alpha} \pm \delta_{\alpha}$ for $\alpha = x,y,z$,
where the $\pm$ signs are used at alternating tetrahedra, so that the breathing factors are each of the form
$1 + 2 \delta_{\alpha}/\tilde{J}_{\alpha}$. In this work we will make the simplifying assumption that all three 
$\delta$'s are equal, so 
the Hamiltonian can be expressed  as 
\begin{equation}
    {\mathcal H}^{\Gamma_{5,6}}_{\rm Breath}={\mathcal H}_{\rm ex}^{\Gamma_{5,6}}+\Delta\mathcal{H},
\label{eq:Hbreath}
\end{equation} where 
\begin{equation}
\Delta\mathcal{H}=\delta\left[\sum_{\langle ij\rangle\in B}\tilde{\bf S}_{i}\cdot\tilde{\bf S}_{j}-\sum_{\langle ij\rangle\in A}\tilde{\bf S}_{i}\cdot\tilde{\bf S}_{j}\right]
\end{equation}
 The general case for where the three $\delta$'s are different can be analyzed following the same procedure presented in this work.

\section{Magnons}

To study the low energy excitations of the system we use the Holstein-Primakoff (HP) transformation\cite{HoPr}
\begin{equation}
\label{eqn:s+}
    \tilde{S}_{i+}=\left(\sqrt{2S-a_{i}^{\dagger}a_{i}}\right)a_{i}, \tilde{S}_{iz}=S-a_{i}^{\dagger}a_{i}
\end{equation}
where $S=1/2$ is the spin quantum number and $a_{i}$ and $a_{i}^{\dagger}$ are bosonic annihilation and creation operators, respectively. The operator defined in Eq.\ \ref{eqn:s+} is cumbersome to deal with due to the square root; hence, we rely on the linear spin wave approximation (LSWA) where $\tilde{S}_{i+}\approx \sqrt{2S} a_{i}$. In magnetic states, the direction for the pseudospin operators is generally not the same as the direction of the physical momentum. For example, in \nd\ the pseudospin operators aligned at an angle $\theta$ with respect to the local $z$-direction defined in Appendix \ref{appendix:a}, which is the direction of the physical momenta in the AIAO ground state.\cite{benton2016,Nd-lhotel2018,Nd-xu2019}

Applying the LSWA to the Hamiltonian in Eq. (\ref{eqn:Gamma56eff}), we obtain a bosonic Hamiltonian 
\begin{equation}
    {\mathcal H}_{\rm ex}^{\Gamma_{5,6}} \approx \frac{3N}{4}\tilde{J}_{z}+{\mathcal H}_{2}
\end{equation}
where $N$ is the total number of magnetic ions and ${\mathcal H}_2$
is quadratic in $a$ and $a^{\dagger}$. Using the Fourier transform of the bosonic operators, the quadratic bosonic Hamiltonian is
\begin{equation}
\label{eqn:H2k}
    {\mathcal H}_{2}=\sum_{\vec{k}}L^{\dagger}_{\vec{k}}M(\vec{k})L_{\vec{k}}
\end{equation}
where
\begin{equation}
L_{\vec{k}}=\left(
\begin{array}{ c c c c l r }
a_{1}(\vec{k}),..,a_{4}(\vec{k}), a^{\dagger}_{1}(-\vec{k}),..,a^{\dagger}_{4}(-\vec{k})
\end{array}\right)^{T},
\label{eq:L}
\end{equation}
and
\begin{equation}
\label{eqn:M}
M(\vec{k})=\left(
\begin{array}{ c c c c c c l r }
\mathcal{J}^{+}R -3\tilde{J}_{z}{\mathbf 1} & \mathcal{J}^{-}R\\
\mathcal{J}^{-}R&\mathcal{J}^{+}R- 3\tilde{J}_{z}{\mathbf 1}\\
\end{array}\right).
\end{equation}
Here ${\mathbf 1}$ is the $4\times 4$ identity matrix, $\mathcal{J}^{\pm}=\frac{\tilde{J}_{x}\pm \tilde{J}_{y}}{2}$, and $R(\vec{k})$ is a $4\times 4$ matrix with
components 
\begin{equation}
    R_{ij}(\vec{k})=\cos\left[\vec{k}\cdot(\vec{r}_{i}-\vec{r}_{j})\right]-\delta_{ij},
    \end{equation}
    where $\vec{r}_{i}$ is the position within a primitive unit cell of the $i$th ($i=1,2,3,4$) magnetic ion.

The next step is to diagonalize ${\mathcal H}_2$  
by finding a 
Bogoliubov transformation of the form $L_{\vec{k}}=Z{\mathcal L}_{\vec{k}}$, where $Z$ is an $8\times 8$ matrix containing the Bogoliubov coefficients and 
\begin{equation}
{\mathcal L}_{\vec{k}}=\left(
\begin{array}{ c c c c l r }
b_{1}(\vec{k}),.., b_{4}(\vec{k}), b^{\dagger}_{1}(-\vec{k}),.., b^{\dagger}_{4}(-\vec{k})\\ 
\end{array}\right)^{T}
\end{equation}
defines a new set of new bosonic operators. The construction of the Bogoliubov matrix $Z$ is discussed in Appendix~\ref{appendix:b}. The diagonalized version of Eq. \ref{eqn:H2k} is 
\begin{equation}
{\mathcal H}_2  =
\sum_{\vec{k}}\sum_{\nu}\varepsilon_{\nu}(\vec{k})
\label{eqn:finalH}
    \left[b_{\nu}^{\dagger}(\vec{k})b_{\nu}(\vec{k})+b_{\nu}^{\dagger}(-\vec{k})b_{\nu}(-\vec{k})\right]
\end{equation}
up to some additive constant. 

For the breathing lattice, we find that to
$M(k)$ (Eq.\ \ref{eqn:M}) we add the matrix 
\begin{equation}
W(\vec{k})=i\delta \left(\begin{array}{cc}
\eta(\vec{k}) & 0 \\
0 & \eta(\vec{k}) \end{array}\right)
\label{eq:W}
\end{equation}
where $\eta(\vec{k})$ is a $4\times 4$ matrix with components
$\eta_{ij}(\vec{k})=\sin\left[\vec{k}\cdot(\vec{r}_{i}-\vec{r}_{j})\right]$. 

In the next section, we apply the procedure presented in the Appendices to analytically determine the Bogoliubov transformation matrix and  magnon dispersions for an AIAO magnetic state with a $\Gamma_{5,6}$ CEF ground state doublet. 
We also examine the effect of a breathing mode on the dispersion. 

\section{Results and Discussion}
\subsection{Energy Dispersion}
\subsubsection{Undistorted Lattice}

We first consider magnons described by the bosonic Hamiltonian 
(\ref{eqn:H2k}) in the absence of any breathing mode. 
According to the results derived in 
Appendix \ref{appendix:R}, the magnon dispersions are 
\begin{equation}
\label{eqn:56generalbands}
     \varepsilon_{\nu}(\vec{k})=\sqrt{(3\tilde{J}_{z}-\tilde{J}_{x}r_{\nu})(3\tilde{J}_{z}-\tilde{J}_{y}r_{\nu})}
\end{equation}
where $r_{1,2} = -1$ and $r_{3,4} = 1 \pm \gamma(\vec{k})$,
with
\begin{eqnarray}
\gamma^2(\vec{k}) &=& 1 + 
\cos( a k_x/2) \cos(a k_y/2) + 
\cos( ak_y/2)\cos(ak_z/2)
 \nonumber \\
& & 
+  \cos(ak_z/2)\cos(ak_x/2),
\label{eq:gamma}
\end{eqnarray}
in agreement with numerical results previously reported.\cite{benton2016,yan2017}
Two bands are doubly degenerate and non-dispersive,
\begin{equation}
\label{eqn:flatbands}
  \varepsilon_{1,2} = \sqrt{(  3\tilde{J}_{z}+\tilde{J}_{x})(3\tilde{J}_{z}+\tilde{J}_{y})}.
\end{equation}

\subsubsection{Breathing lattice}
Here we present the solution to the magnonic Hamilton on a breathing lattice,
\begin{equation}
\label{eqn:Hbreath}
    {\mathcal H}_{\rm breath}=\sum_{\vec{k}}L^{\dagger}_{\vec{k}}(M(\vec{k})+W(\vec{k}))L_{\vec{k}}
\end{equation}
where $L(\vec{k})$, $M(\vec{k})$ and $W(\vec{k})$ are given by
Eqs.\ \ref{eq:L}, \ref{eqn:M} and \ref{eq:W}. Analytic expressions for the band dispersions are derived in Appendix \ref{appendix:Bpy}.
The degenerate non-dispersive bands $\varepsilon_{1,2}$ remain unchanged in the presence of a breathing mode, at least within the LSWA and other assumptions used here. However, the presence of a breathing mode is reflected in the dispersive bands, which are modified to
\begin{equation}
\label{eqn:omegaBPy}
    \varepsilon^2_{3,4}(\vec{k})=b(\vec{k})\pm\sqrt{b^{2}(\vec{k})-c(\vec{k})},
\end{equation}
where
\begin{equation}
\label{eqn:bBPy}
    b(\vec{k})= \sigma(\tilde{J}_x,\tilde{J}_y,\tilde{J}_z,\vec{k})-6 \tilde{J}_z (\tilde{J}_x+\tilde{J}_y)+2 \tilde{J}_x\tilde{J}_y+18\tilde{J}_z^2,
\end{equation}
\begin{equation}
\label{eqn:cBPy}
    c(\vec{k})=\sigma(\tilde{J}_x,\tilde{J}_x, \tilde{J}_z,\vec{k})\sigma(\tilde{J}_y,\tilde{J}_y,\tilde{J}_z,\vec{k}),
\end{equation}
and
\begin{eqnarray}
    \sigma(J_x,J_y,J_z,\vec{k})& = &\left(\gamma^2(\vec{k})-1\right) \left(J_xJ_y-\delta ^2\right)
    \nonumber \\
   & &  +3 \left(\delta ^2+(J_x+J_y) J_z-3 J_z^2\right).
   \label{eq:sigma}
\end{eqnarray}

\begin{figure}[ht]
    \includegraphics[width=0.45\textwidth]{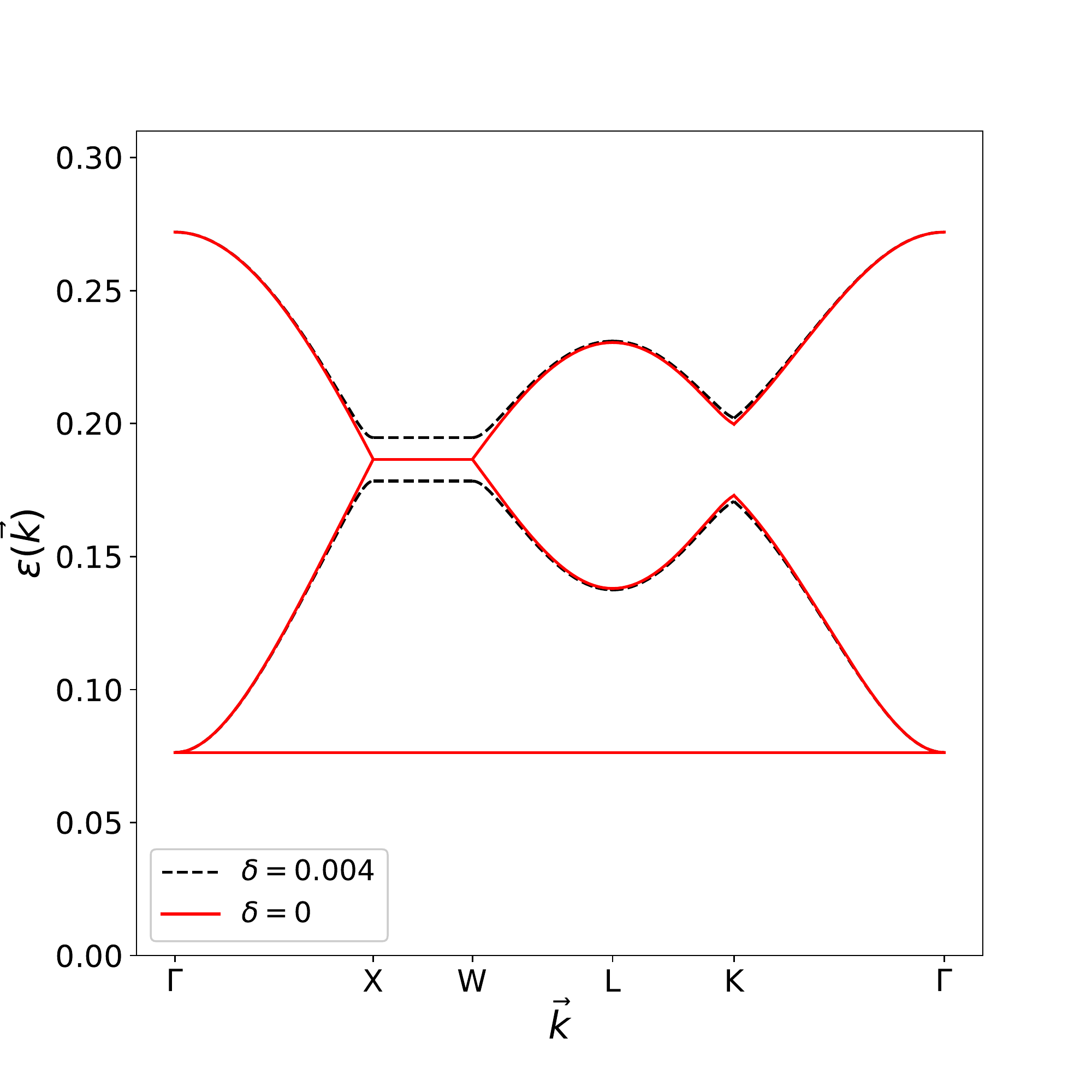}
    \caption{Magnon dispersions computed using  the exchange constants of \nd\ (discussed in the text) with $\delta=0$ (red solid) and $\delta = 0.004$ meV (black dashed).
    }
    \label{fig:BPy}
\end{figure}

As an example, we compute the magnon spectrum using the exchange constants of ${\rm Nd_{2}Zr_{2}O_{7}}$ which were reported  in Ref.\ \onlinecite{Nd-xu2019}.  In Fig.\ \ref{fig:BPy} we plot the magnon 
energies using $\delta=0$ (no distortion) and $\delta = 0.004$ meV. We find that a breathing mode distortion is reflected on the $k$-space path between the $X$ and $W$ points, where the degeneracy of the top band is lifted, resulting in a gap between the two upper bands, similar to  the findings in Ref.\ \onlinecite{Essafi_2017}. The gap $\Delta$ between the upper bands along the path $X\to W$  is linear in $\delta$,
\begin{equation}
\label{delgenlinfit}
     \Delta\approx\frac{2(\tilde{J}_{x}+\tilde{J}_{y}-6\tilde{J}_{z})}{\sqrt{(\tilde{J}_{x}-3\tilde{J}_{z})(\tilde{J}_{y}-3\tilde{J}_{z})}}\delta.
\end{equation}

\subsection{The dynamical structure factor}

\begin{figure*}[ht]
\subfloat{\includegraphics[width = 6.8in]{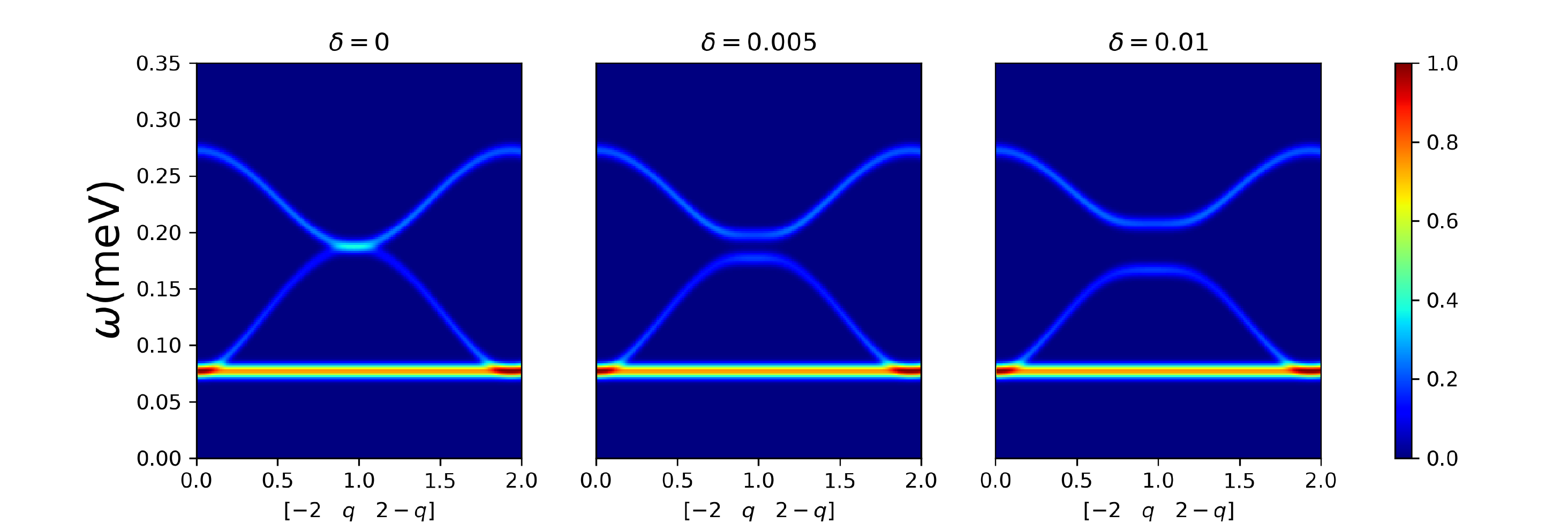}}
\captionsetup{justification   = raggedright,
              singlelinecheck = false}
\caption{The dynamical structure factor 
along the 
direction computed using   $\tilde{J}_{x}=0.091$ meV, $\tilde{J}_{y}=0.014$ meV, and $\tilde{J}_{z}=-0.046$ meV, $\theta=0.98$ rad, and lattice distortion $\delta = 0$ meV (left), $\delta = 0.005$ meV (middle), and $\delta = 0.010$ meV (right). 
We set the temperature to 0 K
and a Gaussian standard deviation of 0.004.
}
\label{fig:Sspectrum}
\end{figure*}

An important quantity that is experimentally accessible through inelastic neutron scattering is the dynamical structure factor,\cite{yan2017}
\begin{eqnarray}
\label{eqn:Sq}
    \mathcal{S}(\vec{q},\omega)&=&\int dt e^{-i\omega t}    \sum_{\alpha\beta}\left(\delta_{\alpha\beta}-\frac{q_{\alpha}q_{\beta}}{q^{2}}\right)\nonumber \\
   & & \times \sum_{i,j=1}^{4}\langle \hat{J}_{i}^{\alpha}(-\vec{q},0)\hat{J}_{j}^{\beta}(\vec{q},t)\rangle
\end{eqnarray}
where $i=1,2,3,4$ is the site number within the primitive unit cell, and $\alpha = x,y,z$ refer to global axes. Eqn.\ \ref{eqn:Sq} can be expressed in terms of the momenta for local axes  $\hat{J}_{i\alpha}$ using the definitions in Appendix \ref{appendix:a}, however in the restriction to the  dipolar-octupolar ($\Gamma_{5,6}$) CEF doublet, the only non-vanishing component of the physical angular momentum is $\hat{J}_{iz}$. 
Finally, these are   related to the pseudo-spin operators introduced in Eqs.\ \ref{eqn:Gamma56} and \ref{eqn:Gamma56eff}.
In the LSWA, only the transverse components  components (perpendicular to $\tilde{S}_z$) contribute to the spin-spin correlation function. Our analytic form of the Bogoluibov transformation (contructed in Appendices \ref{appendix:gamma56} and \ref{appendix:Bpy}) is used to express the spin-spin correlation function in terms of the normal modes (magnons). The dynamical structure factor is shown in Fig.\ \ref{fig:Sspectrum} for different values of the lattice distortion.

Fig. \ref{fig:Sspectrum} shows the intensity map of the dynamical structure factor along the line $[-2 \,\, q \,\, 2-q]$.  The most pronounced changes to the magnon dispersions occur in the vicinity of  $[-2 \,\, 1 \,\,1]$, which is equivalent to the $X$-point of the first Brillioun zone. Here in the presence of a breathing mode the degeneracy is lifted and there is a distinct flattening of the bands. 

\section{Conclusions} 
 
In this work we presented the exact analytic diagonalization of the magnon Hamiltonian for magnetic pyrochlores in the linear spin wave approximation, applied to the special case  where the magnetic ions have a $\Gamma_{5,6}$ CEF ground state doublet with an ordered  AIAO magnetic ground state.\cite{benton2016,yan2017} As an illustration, we  applied our findings to \nd\ for which the exchange constants and the CEF states are known.\cite{benton2016,Nd-lhotel2018,Nd-xu2019} We also considered the breathing pyrochlore lattice and found the analytic form of the energy dispersion for all $\vec{k}$. In the special case we considered, the signature of a lattice distortion is the degeneracy-lifting of the upper magnon band between the $X$ and $W$ points in $k$-space. 

\begin{acknowledgments}
This work was supported by NSERC of Canada.
\end{acknowledgments}

\appendix

\section{The local reference frame}
\label{appendix:a}

The primitive unit cell of the RE system is a single tetrahedron with the four rare earth ions at the positions $(5/8,5/8,5/8)$, $(3/8,3/8,5/8)$, $(3/8,5/8,3/8)$ and $(5/8,3/8,3/8)$ for site numbers 1, 2, 3, and 4 respectively.
We define a local coordinate system such that the local $z$-axes are the 
the $C_3$ axes for each site: 
\begin{eqnarray*}
\hat{x}_{1}=(1,1,-2)/\sqrt6  &  \hspace{.2in} & \hat{x}_{2}=(-1,-1,-2)/\sqrt6 \\
 \hat{y}_{1}=(-1,1,0)/\sqrt2 & & \hat{y}_{2}=(1,-1,0)/\sqrt2  \\
 \hat{z}_{1}=(1,1,1)/\sqrt3 & &  \hat{z}_{2}=(-1,-1,1)/\sqrt3 \\
 \hat{x}_{3}=(-1,1,2)/\sqrt 6& & \hat{x}_{4}=(1,-1,2)/\sqrt6  \\
\hat{y}_{3}=(1,1,0)/\sqrt2 & & \hat{y}_{4}=(-1,-1,0)/\sqrt2\\
 \hat{z}_{3}=(-1,1,-1)/\sqrt3 & &  \hat{z}_{4}=(1,-1,-1)/\sqrt3
\end{eqnarray*}

\section{The Bogoliubov transformation matrix}
\label{appendix:b}
The main task here is to find the Bogoliubov transformation matrix $Z$ which satisfies the following conditions simultaneously:
\begin{eqnarray}
    Z^{-1}{\mathcal G}MZ& = & \left(
\begin{array}{cc}
{\mathcal E} & 0\\
0&-{\mathcal E}\\
\end{array}\right)
\label{eqn:BT1}
 \\
    Z{\mathcal G}Z^{\dagger}& = & {\mathcal G}=\left(
\begin{array}{ c c c c c c l r }
{\mathbf 1} & \mathbf{0}\\
\mathbf{0}&-{\mathbf 1}\\
\end{array}\right),
\label{eqn:BT2} 
\\
    Z^{\dagger}MZ& = & \left(
\begin{array}{cc}
{\mathcal E} & 0\\
0&{\mathcal E}\\
\end{array}\right)
\label{eqn:BT3}
\end{eqnarray}
where $M$ is the $8\times 8$ matrix given by Eq.\ \ref{eqn:M}.
We assume that ${\mathcal G}M$ has $q$ distinct eigenvalues (where $q\leq 8$), each with degeneracy $d_{i}$.\cite{Demastro04} 
We begin by finding the normalized eigenvectors of ${\mathcal G}M$ and arranging their
components in columns, such that the eigenvectors belonging to the same eigenvalue 
are next to each other.
We call this matrix $\tilde{Z}$. Next, we assume that there is a block-diagonal matrix $P$, such
that $Z=\tilde{Z}P$, where the sizes of the blocks of $P$ are $d_{i}\times d_{i}$. Considering Eq.\ \ref{eqn:BT2} and that ${\mathcal G}^{-1} = {\mathcal G}$, we have\cite{Demastro04}
\begin{equation}
    P{\mathcal G}P^{\dagger}=(\tilde{Z}^{\dagger}{\mathcal G}\tilde{Z})^{-1}.
\end{equation}
By construction, the matrix $W=(\tilde{Z}^{\dagger}{\mathcal G}\tilde{Z})^{-1}$ is Hermitian and block-diagonal, with $q$ blocks with dimensions $d_{i}\times d_{i}$. 
For each block we write 
\begin{equation}
\label{eqn:pw}
    \pm P_{i}P_{i}^{\dagger}=W_{i}
\end{equation}
where the sign 
is positive for blocks in the upper half of $P$ and negative otherwise. Since each block $W_{i}$ is Hermitian, each $W_i$ can be diagonalized by a unitary matrix
$X_i$,  
\begin{equation}
    W_{i}=X_{i}D_{i}X_{i}^{-1}
\end{equation}
where $D_{i}$ is a diagonal matrix containing the eigenvalues of $W_{i}$. Consequently, the solution of $P_{i}$ is\cite{Demastro04}
\begin{equation}
    P_{i}=X_{i}\sqrt{\pm D_{i}}X_{i}^{-1}.
\end{equation}
Thus, we can construct the Bogoliubov transformation matrix $Z = \tilde{Z} P $.

\section{Eigenvalues and eigenvectors of $R$ and~$\eta$}
\label{appendix:R}
In this appendix we present analytic forms for the eigenvalues and eigenvectors of 
the matrix $R$ (see Eq. \ref{eqn:M}). The results will be used in Appendix \ref{appendix:gamma56} to 
find the analytic form of the magnon dispersions and to construct the Bogliubov transformation.
To shorten the notation we use $\{\cos(ak/4),\sin(ak/4)\}=\{c(k),s(k)\}$. The eigenvalues of $R$ are
\begin{equation}
\label{eqn:eigensR}
    r_{1}=r_{2}= -1, \; \; \; r_{3,4}\equiv r_{\pm}=1\pm \gamma
\end{equation}
where   $\gamma^{2}=1+c(2k_x)c(2k_y) + c(2k_y)c(2k_z)
    +c(2k_z)c(2k_x)$ (as in Eq.\ \ref{eq:gamma}). The eigenvectors of $R$ corresponding to $r_{1,2}=-1$ are
\begin{equation}
\label{eqn:Ru12}
    \Bigg\{\left[
\begin{array}{cccc}
    s\left(k_x-k_z\right)\\
    s\left(k_y+k_z\right)\\
    0\\
    -s\left(k_x+k_y\right)\\\end{array}\right], \left[
\begin{array}{cccc}
    s\left(k_y-k_z\right)\\
    s\left(k_x+k_z\right)\\
    -s\left(k_x+k_y\right)\\
    0\\\end{array}\right]\Bigg\}.
\end{equation}
Note that these are not orthogonal as they correspond to the same eigenvalue. However, orthogonal eigenvectors can be easily found as simple linear combinations. We call the orthonormalized eigenvectors $\vec{v}_1$ and $\vec{v}_2$.

We write the other two eigenvectors (corresponding to
$r_{3,4}$) as
\begin{equation}
\label{eqn:upm}
    \vec{u}_{3,4} = \vec{u}_{\pm} = \left[
\begin{array}{cccc}
f_{1}^{\pm}, f_{2}^{\pm}, f_{3}^{\pm},f_{4}^{\pm}
\end{array}\right]^{T}
\end{equation}
where
\begin{equation}
    f_{n}^{\pm}(\vec{k})=g_{n}(\vec{k})+r_\pm c(\vec{k}\cdot\vec{r}_{n})
\end{equation}
and $\vec{r}_{n}$ is the position of the $n$th rare-earth ion within the primitive cell. The functions $g_{n}(\vec{k})$ are
\begin{equation}
    g_{n}(\vec{k})= 
     \begin{cases}
       \phi(k_x,k_y-k_z,k_y+k_z) &\quad n=1\\
      -\phi(k_z,k_y-k_x,\pi/2)&\quad n=2\\
      -\phi(-k_y,k_z-k_x,\pi/2)&\quad n=3\\
      -\phi(-k_x,k_y-k_z,\pi/2)&\quad n=4\\
     \end{cases}
\end{equation}
where
\begin{equation}
    \phi(x,y,z)=c(2x+z)c(y)+c^{2}(z).
\end{equation}
The normalized eigenvectors are $\vec{v}_{\pm}=\frac{\vec{u}_{\pm}}{||\vec{u}_{\pm}||}$. 
One can easily verify that $\vec{v}_{\pm}\cdot\vec{v}_{1,2}=0$ for all $\vec{k}$. The orthonormalized version of this set of eigenvectors is needed for the analytic calculation of the Bogoliubov matrix $Z$, as discussed in Appendices \ref{appendix:gamma56} and \ref{appendix:Bpy}.\\

$R$ and $\eta$ share the eigenvectors $\vec{v}_{1,2}$  (Eq.\ \ref{eqn:Ru12}) and the corresponding eigenvalues for $\eta$ are $0$ (doubly degenerate), {\em i.e.} $\eta \vec{v}_{1,2} = 0$.  To find the other two eigenvectors of 
$\eta$ we consider the following. One can easily verify the relations between $\eta$ and $R$:
\begin{equation}
    \{R-\boldsymbol{1},\eta\}=0,\quad \eta^{2}=R^{2}-2R-3.
\end{equation}
Also, the action of $\eta$ on $\vec{v}_{\pm}$
is
\begin{equation}
    \eta\vec{v}_{\pm}=\sqrt{\gamma^{2}-4}\vec{v}_{\mp}.
\end{equation}
Consequently, the other two
eigenvectors of $\eta$ are
\begin{equation}
 \vec{z}_{3,4} =  \vec{v}_{+}\pm\vec{v}_{-}
\end{equation}
with eigenvalues  $\pm\sqrt{\gamma^{2}-4}$,
 which are generally complex away from the $\Gamma$ point. 

\section{Analytic diagonalization of the magnonic Hamiltonian for an undistorted lattice}
\label{appendix:gamma56}
We can 
 write the matrix ${\mathcal G}M$  in the form
\begin{equation}
    {\mathcal G}M(k)=\left(
\begin{array}{ c c c c c c l r }
\mathcal{A} & \mathcal{B}\\
-\mathcal{B}&-\mathcal{A}\\
\end{array}\right)
\label{eq:D1}
\end{equation}
where $\mathcal{A} = {\cal J}^{+} R- 3 \tilde {J}_{z} {\bf 1}$,
 $\mathcal{B} = {\cal J}^{-} R$,
 and $R$ and ${\cal J}^{\pm}$ are defined after
 Eq.\ \ref{eqn:M}.
The eigenvalue equation for a matrix of this form (with $\mathcal{A}$ and $\mathcal{B}$ commuting matrices) reduces to the secular determinant\cite{blocks} 
\begin{equation}
\label{eqn:ABR}
    |(\mathcal{A}^{2}-\mathcal{B}^{2})-\varepsilon^{2}\mathbf{1}|=(\tilde{J}_{x}\tilde{J}_{y})^{4}|R^{2}-aR+b\mathbf{1}|=0,
\end{equation}
where $a=\frac{3\tilde{J}_{z}(\tilde{J}_{x}+\tilde{J}_{y})}{\tilde{J}_{x}\tilde{J}_{y}}$ and $b=\frac{(9\tilde{J}_{z}^{2}-\varepsilon^{2})}{\tilde{J}_{x}\tilde{J}_{y}}$.  This   reduces to $b=
r(a-r)$, where $r$ are the eigenvalues of $R$. Solving for $\varepsilon$, we obtain the general form of the magnon dispersions  given in Eq.\ \ref{eqn:56generalbands}. 

To find the full Bogoliubov transformation, we first apply the unitary transformation 
\begin{equation}
\label{eqn:unitary}
    {\mathcal U}=\frac{1}{\sqrt{2}}\left(
\begin{array}{ c c c c c c l r }
\mathbf{1}&\mathbf{1}\\
-\mathbf{1}&\mathbf{1}\\
\end{array}\right)
\end{equation}
to  ${\mathcal G}M(k)$.  We obtain
\begin{equation}
\label{eqn:GM561}
    {\mathcal U}{\mathcal G}M{\mathcal U}^{\dagger}=\left(\begin{array}{cc}
    \mathbf{0}&-(\mathcal{A}-\mathcal{B})\\
    -(\mathcal{A}+\mathcal{B})&\mathbf{0}\\
    \end{array}\right).
\end{equation}
Now, one can easily verify that the normalized eigenvectors of the $8\times 8$ matrix  ${\mathcal U}{\mathcal G}M{\mathcal U}^{\dagger}$ are 
\begin{equation}
    \vec{\tilde{X}}^{\pm}_{i}=\left[1+\left(\frac{-\tilde{J}_{x}r_{i}+3\tilde{J}_{z}}{\varepsilon_{i}}\right)^{2}\right]^{-\frac{1}{2}}\Bigg(
\begin{array}{ c c c c c c l r }
\vec{v}_{i}\\
\pm\frac{-\tilde{J}_{x}r_{i}+3\tilde{J}_{z}}{\varepsilon_{i}}\vec{v}_{i}\\
\end{array}\Bigg),
\end{equation}
where $\vec{v}_{i}$ are the orthonormal eigenvectors of the matrix $R$ (see Eqns.\ \ref{eqn:Ru12} and \ref{eqn:upm}). Consequently, using the unitary matrix ${\mathcal U}$, we find the normalized eigenvectors of ${\mathcal G}M$ to be
\begin{equation}
    \vec{X}_{i}^{\pm}=  {\mathcal U}^{\dagger}\vec{\tilde{X}}^{\pm}_{i}=
    \left(
\begin{array}{ c c c c c c l r }
\vec{w}_{i}^{\mp}\\
\vec{w}_{i}^{\pm}\\
\end{array}\right)
\label{eqn:xgencase}
\end{equation}
where 
$$\vec{w}_{i}^{\pm}=\frac{1\pm\frac{-\tilde{J}_{x}r_{i}+3\tilde{J}_{z}}{\varepsilon_{i}}}{\sqrt{2+2\left(\frac{-\tilde{J}_{x}r_{i}+3\tilde{J}_{z}}{\varepsilon_{i}}\right)^{2}}}\vec{v}_{i}.$$ 
Next, we
use the eigenvectors of ${\mathcal G}M$ to 
 construct the matrix $\tilde{Z}$ (see Appendix \ref{appendix:b}). Lastly, to find the Bogoliubov matrix $Z$, we follow the procedure described in Appendix \ref{appendix:b} by which we find the matrix $P$ and calculate $Z=\tilde{Z}P$. The matrix ${\mathcal G}M$ has 6 unique eigenvalues of which two are doubly degenerate (corresponding to the flat bands). Thus, the matrix $P$ has six blocks where two of them are $2\times 2$ diagonal blocks and the rest are $1\times 1$. 
Thus, $P$ is a diagonal matrix of the form $P={\rm diag}[p_{1},..,p_{4},p_{1},..p_{4}]$, where
\begin{equation}
\label{eqn:pgencase}
    p_{i}=\frac{1}{2}\sqrt[4]{
    \frac{3\tilde{J}_{z}-\tilde{J}_{y}r_{i}}{3\tilde{J}_{z}-\tilde{J}_{x}r_{i}}}
    \sqrt{2+2\left(\frac{-\tilde{J}_{x}r_{i}+3\tilde{J}_{z}}{\varepsilon_{i}}\right)^{2}}.
\end{equation}
Defining $\vec{z}_{i}^{\pm}=p_{i}\vec{w}_{i}^{\pm}$, and
$Q_{\pm}=\left(\begin{array}{cccc}
\vec{z}_{1}^{\pm} & \vec{z}_{2}^{\pm} & \vec{z}_{3}^{\pm} & \vec{z}_{4}^{\pm}
\end{array}\right)$, we write the Bogoliubov matrix in block form as 
\begin{equation}
    Z=\left(\begin{array}{cc}
        Q_{-} & Q_{+} \\
        Q_{+} & Q_{-}\\ 
    \end{array}\right)
    \label{eqn:z1stcase}
\end{equation}
which satisfies Eqs. (\ref{eqn:BT1}-\ref{eqn:BT3}).

\section{Analytic Diagonalization for breathing lattice} \label{appendix:Bpy}
Following the same procedure in the previous appendix, we find that
\begin{equation}
\label{eqn:UGMUBPy}
    {\mathcal U}{\mathcal G}M{\mathcal U}^{\dagger}=\left(\begin{array}{cc}
    \mathbf{0}&-(\mathcal{A}+i\delta\eta-\mathcal{B})\\
    -(\mathcal{A}+i\delta\eta+\mathcal{B})&\mathbf{0}\\
    \end{array}\right).
\end{equation}
The eigenvalue equation $|{\mathcal U}{\mathcal G}M{\mathcal U}^{\dagger}-\varepsilon|=0$ reduces to the following $4\times 4$ secular determinant\cite{blocks}
\begin{equation}
    |({\mathcal A}+i\delta \eta(k))^{2}-{\mathcal B}^{2}\pm i\delta\mathcal{J}^{-}[R,\eta]-\varepsilon^{2}(k)|=0,
\end{equation}
where the $\pm$ follows from the properties of Schur determinants.\cite{blocks} Without loss of generality, we will consider the $+$ sign and solve the eigenvalue equation 
$|\Lambda-\varepsilon^{2}(k)|=0$, where $\Lambda=({\mathcal A}+i\delta \eta(k))^{2}-{\mathcal B}^{2}+i\delta\mathcal{J}^{-}[R,\eta]$. The general form of the eigenvectors is a linear 
combination of the orthonormalized eigenvectors of $R$:
\begin{equation}
    \vec{y}=c_{1}\vec{v}_{1}+c_{2}\vec{v}_{2}+c_{3}\vec{v}_{3}+c_{4}\vec{v}_{4}.
\end{equation}
Defining $\vec{c}=(c_{1},c_{2},c_{3},c_{4})^{T}$, we map the eigenvalue equation to one using the eigenvectors of $R$ as  the basis vectors:
\begin{equation}
\label{eqn:blockformBpy}
    \left(\begin{array}{cc}
    \Omega_{1}&\boldsymbol{0}\\
    \boldsymbol{0}&\Omega_{2}\\
    \end{array}\right)\vec{c}=\varepsilon^{2}\vec{c}
\end{equation}
where $\Omega_1$ and $\Omega_2$ are $2\times 2$ matrices,
\begin{eqnarray}
    \Omega_{1}&=&\left[(\mathcal{J}^{+}+3\tilde{J}_{z})^{2}-(\mathcal{J}^{-})^{2}\right]\boldsymbol{1},
\\
    \Omega_{2}&=&\left(\begin{array}{cc}
    a_{-}^{2}-\delta^{2}t^{2}-b_{-}^{2}&i\delta t\left[a_{+}+a_{-}-2\gamma\mathcal{J}^{-}\right]\\
    i\delta t\left[a_{+}+a_{-}+2\gamma\mathcal{J}^{-}\right]&a_{+}^{2}-\delta^{2}t^{2}-b_{+}^{2}
    \end{array}\right) \nonumber \\
    & & 
\end{eqnarray}
and $a_{\pm}=\mathcal{J}^{+}r_{\pm}-3\tilde{J}_z$, $b_{\pm}=\mathcal{J}^{-}r_{\pm}$, and $t=\sqrt{\gamma^{2}-4}$. The eigenvalues of $\Omega_{1}$ are $(\mathcal{J}^{+}+3\tilde{J}_{z})^{2}-(\mathcal{J}^{-})^{2}$ (doubly degenerate) producing the flat bands in Eq.\ \ref{eqn:flatbands}, and the eigenvalues of $\Omega_{2}$ are the dispersive bands energies (squared) given by
Eqs.\ \ref{eqn:omegaBPy} - \ref{eq:sigma}.

Next, we calculate the Bogoliubov matrix $Z$. Let $X=\left(\begin{array}{cc}
x_{11}&x_{12}\\
x_{21}&x_{22}\\
\end{array}\right)$ be a $2\times 2$ matrix that diagonalizes $\Omega_{2}$; that is, $\Omega_{2}=XDX^{-1}$, where $D$ is a diagonal matrix containing the eigenvalues of $\Omega_{2}$. For a $2\times 2$ matrix $X$ is easily found. Then the matrix that diagonalizes the $4\times 4$ matrix in Eq. \ref{eqn:blockformBpy} is 
\begin{equation}
    W=\left(\begin{array}{cc}
     \boldsymbol{1}&\boldsymbol{0}\\
     \boldsymbol{0}&X\\
    \end{array}\right).
\end{equation}
Thus, we conclude that the four eigenvectors of $\Lambda$ are
\begin{eqnarray}
    \vec{y}_{1}=\vec{v}_{1},  &\quad& \vec{y}_{2}=\vec{v}_{2}, \\
    \vec{y}_{3}=x_{11}\vec{v}_{3}+x_{21}\vec{v}_{4}, &\quad {\rm and}\quad& \vec{y}_{4}=x_{12}\vec{v}_{3}+x_{22}\vec{v}_{4}.
\end{eqnarray}
Using this information, we find that the eigenvectors of ${\mathcal G}M$ are of the form 
\begin{equation}
    \vec{X}^{\pm}_{i}=\frac{1}{\sqrt{2}}\left(\begin{array}{cc}
    \pm\varepsilon_{i}^{-1}{\mathcal C} & -\mathbf{1}\\
    \pm\varepsilon_{i}^{-1}{\mathcal C} &\mathbf{1}\\ 
    \end{array}\right)\left(\begin{array}{cc}
    \vec{y}_{i}\\
    \vec{y}_{i}\\
    \end{array}\right),
\end{equation}
where ${\mathcal C}=-({\mathcal A}+i\delta\eta-{\mathcal B})$. Using these vectors, we can determine $\tilde{Z}=\left(\begin{array}{ccccccccccccc}
\vec{X}^{+}_{1}&\cdots & \vec{X}^{+}_{4}&\vec{X}^{-}_{1}&\cdots & \vec{X}^{-}_{4}
\end{array}\right)$, which we then use to find the matrix
\begin{equation}
\tilde{Z}^{\dagger}{\mathcal G}\tilde{Z}=\left(\begin{array}{cc}
F^{(1)}&F^{(2)}\\
F^{(3)}&F^{(4)}\\
\end{array}\right),
\end{equation}
where $F^{(1)}_{ij}=-(\varepsilon_{i}^{-1}+\varepsilon_{j}^{-1})\vec{y}_{i}\cdot {\mathcal C}\vec{y}_{j}$, $F^{(2)}_{ij}=-(\varepsilon_{i}^{-1}-\varepsilon_{j}^{-1})\vec{y}_{i}\cdot {\mathcal C}\vec{y}_{j}=0= F^{(3)}_{ij}$ and $F^{(4)}_{ij}=(\varepsilon_{i}^{-1}+\varepsilon_{j}^{-1})\vec{y}_{i}\cdot {\mathcal C}\vec{y}_{j}=-F^{(1)}_{ij}$. Using the relation $W^{-1}=\tilde{Z}^{\dagger}{\mathcal G}\tilde{Z}$ together with Eq. \ref{eqn:pw}, we find $P={\rm diag}[p_{1},..,p_{4},p_{1},..p_{4}]$, where
\begin{equation}
    p_{i}=\sqrt{-\frac{\varepsilon_{i}}{2\vec{y}_{i}\cdot {\mathcal C}\vec{y}_{i}}}.
\end{equation}
Finally the Bogoliubov matrix for the breathing lattice is $Z=\tilde{Z}P$.

\nocite{*}
\bibliography{magnons}

\end{document}